# Estimating the Porosity Structure of Granular Bodies Using the Lane–Emden Equation Applied to Laboratory Measurements of the Pressure–Density Relation of Fluffy Granular Samples

Tomomi Omura and Akiko M. Nakamura

Department of Planetology, Kobe University, 1-1 Rokkodai-cho, Nada-ku, Kobe, Hyogo, 657-8501, Japan

Send editorial correspondence to:
Tomomi Omura
Department of planetology
Kobe University
1-1 Rokkodai-cho, Nada-ku
Kobe, 657-8501
Japan
Email address: tomura@stu.kobe-u.ac.jp

**Abstract**

The porosity structure of a granular body is an important characteristic that affects evolutionary changes in the body. We conducted compression experiments using fluffy granular samples with various particle sizes, shapes, and compositions. We approximated the pressure-filling factor relationship of each sample with a power law (a modified polytropic relationship). We also fit our previous data and literature data for fluffy granular samples using a power-law equation. The fitting with a power-law form was as good as that achieved with the equations used for powders in previous studies. The polytropic indices obtained in the current study ranged from ~0.01 to ~0.3 and tended to decrease with increasing particle size for samples of similar porosities. We calculated the radial porosity structure and bulk porosity of granular bodies with various radii using the Lane–Emden equation. The results provide the initial, most porous structures of accreted primordial bodies, or re-accumulated rubble-pile bodies consisting of particles that have compression properties similar to those of the assumed granular materials. A range of porosity structures is allowed for a body of given size and macroporosity, depending on the compression properties of the constituent granular material.

Key words: comets: general — methods: laboratory: solid state — minor planets, asteroids: general— planets and satellites: formation

## 1. INTRODUCTION

Granular objects, such as planetesimals consisting of dust grains and rubble-pile bodies consisting of impact debris, have been ubiquitous throughout the evolution of the solar system (Davis et al. 1979; Weidenschilling 1980; Fujiwara et al. 2006; Kataoka et al., 2013b). The porosity of these objects is an important characteristic that affects their response to impacts as well as their thermal properties. Therefore, the porosity structure determines the evolutionary path of an object. For instance, laboratory and numerical simulations of impacts with small bodies with different internal structures have shown that the damaged or fractured region remains localized when the body has a porous internal structure (Love et al. 1993; Asphaug et al. 1998). The thermal evolution of meteorite parent bodies, coated with a thick, insulating, porous regolith accumulated through repeated impacts, and the re-accumulation of ejecta on the surface of bodies with a primordial porous granular structure, has been shown to differ greatly from that of simple rocky bodies (Akiridge et al. 1998; Henke et al. 2012).

Recently, there has been an increase in data on the bulk density of small bodies (Consolmagno et al. 2008; A'Hearn 2011; Baer et al. 2011), providing some indication of internal density structure. A range of density structures should arise from soil pressure due to the equilibrium of self-gravity, centrifugal force, tidal force, the presence of rocks, compression by impacts, and compaction due to impact-induced vibrations. In particular, the density structure due to soil pressure provides the initial, most porous structure when the centrifugal and tidal forces can be neglected. Previous studies estimated bulk density corresponding to the pressure due to self-gravity based on compression curves obtained in laboratory experiments (Yasui & Arakawa 2009) or numerical simulations (Kataoka et al. 2013b), where uniform pressure and density were assumed and a density gradient was not taken into account.

In this study, we conducted compression experiments with fluffy samples having various initial porosities, using particles with various sizes, shapes, and compositions. Power-law equations were fit to the compression curves and the Lane–Emden equation was then applied, even though this equation was originally intended for fluids, to calculate the density structure of granular bodies with different radii, assuming that the density structure is determined by soil pressure due only to self-gravity in the vertical direction.

## 2. COMPRESSION EXPERIMENTS

### 2.1 Experimental Procedure

We used irregular alumina particles, polydisperse spherical silica particles, and polydisperse irregular silica sand with three different sized grains as sample powders. Table 1 lists the physical properties of the sample powders. The silica sand grains were used in a previous study (Omura et al. 2016), but we updated the density of the grain according to a measurement of true density (2.645 g $cm^{-3}$) made using a pycnometer. We mimicked the porosity of fluffy granular bodies using these materials. The compression properties of a granular body may vary depending on the structure of the layer (i.e., porosity) and the physical properties of constituent particles. The relationship between the characteristics of the powder and the compression properties of samples is discussed in Section 2.3.

The cumulative volume fractions of the sample were determined using a laser diffractometer (SHIMAZU SALD-3000S) installed at Kobe University and are shown in Fig. 1. The cumulative volume fraction is defined as $F_{vol}$ (< x) = $d_x$. Table 1 lists the median diameter ($d_{50}$) and ratio $d_{85}/d_{15}$ of each sample. The ratio $d_{85}/d_{15}$ is an indicator of the broadness of the size distribution: the larger the ratio, the broader the size distribution. Note that the measured sample powders may have contained lumps; therefore, the particle size distribution shown in Fig. 1 does not represent the individual particle size distribution, especially in the case of alumina particles. These lumps may be similar in structure to the dust aggregate used in a previous study (Weidling et al. 2012) and cometary dust particles (Bentley et al. 2016). For this sample, we show both median diameter obtained by measurement and manufacturer's information about particle diameter. We sieved the samples using a 500 μm mesh screen into a cylindrical container and then used a spatula to level off the part of the bed that exceeded the height of the container. The initial filling factor of the samples was adjusted by tapping and piston loading.

In our previous study, granular samples were compressed by a centrifuge to simulate the body force of soil pressure due to self-gravity inside small bodies (Omura et al. 2016). The results of centrifuge compression are consistent with those of piston compression when the density gradient inside the centrifuge sample is taken into consideration (Suzuki et al. 2004). Therefore, in the current study, we performed piston compression to obtain the compression curves of the samples. The samples were loaded by a piston fixed to a compressive testing machine installed at Kobe University (EZ-Graph). The uniaxial pressure on the surface of a sample ranged from 1,000 to $4 \times 10^5$ Pa for the alumina (0.1 μm) and silica beads (1.7 μm), and from $1.5 \times 10^4$ to $4 \times 10^5$ Pa for the silica sand. The pressure acting on the sample was calculated by dividing the compressive force by the cross-sectional area of the piston. In fact, friction between the particles and the container wall produces vertical pressure acting on the bottom of the sample that is smaller than that acting on the surface (Janssen effect; e.g. Duran 2000).

We calculated the pressure on the bottom of the sample in the same way as described in Omura & Nakamura (2017), and confirmed that the ratio of the vertical pressures acting on the bottom and surface of a sample was between 0.75 and 0.93. The bulk density of the samples was calculated by the volume and mass of the sample, and the volume of the sample was estimated from its height, which was calculated using the height of the piston. The details of the experimental procedure are the same as described in Omura & Nakamura (2017) for alumina (0.1 μm) and silica beads (1.7 μm), and in Omura et al. (2016) for silica sand.

## 2.2 Experimental Results

Figure 2 shows the results of the compression experiments. The vertical axis shows the applied pressure ($P$) normalized by the tensile strength of the sample particles ($Y$). We adopted $Y = 6.0 \times 10^7$ Pa for $SiO_2$ (Kaye & Laby 2005) and $Y = 2.8 \times 10^8$ Pa for $Al_2O_3$ (manufacturer's information). We took tensile strength to be the strength of the sample grains, because tensile strength and crushing strength have been shown to be almost the same in laboratory measurements (S. Shigaki 2016, private communication). Here, we neglected the effects of size on particle strength, which increases with decreasing particle size (Yashima & Saito 1979). The horizontal axis is the filling factor of the sample:

$$f = \rho_{bulk}/\delta \quad (1)$$

where $\rho_{bulk}$ and δ are the bulk density and grain density of the sample, respectively (listed in Table 1). For alumina and silica beads, the average values of three experiments are shown. The vertical and horizontal error bars correspond to the standard deviation of the averaged pressure and the uncertainty of the filling factor, respectively. For silica sand, the result of a single measurement with an error bar corresponding to the uncertainty in sample volume is shown (see Omura et al., 2016; Omura & Nakamura 2017). The filling factor for the samples remained almost constant at its initial value until the threshold pressure, defined as the yield strength, was reached, at which point compaction started (Omura & Nakamura. 2017). Here, we show only compression curves with a pressure range larger than the yield strength. For silica sand, no such region existed and all data are shown in Fig. 2. The density change of the largest silica sand particles (73 µm) due to the applied pressure was smaller than that of the 13 µm and 19 µm samples; in other words, that particular sample was the least compressible, probably because it had the narrowest size distribution. The silica sand (19 µm) with the broadest size distribution had the highest density, because void spaces between large particles can be filled effectively by small grains.

## 2.3 Fitting the Results

The polytropic relationship used in the Lane–Emden equation (e.g. Chandrasekhar, 1957) is as follows:

$$P = K\rho^{\frac{n+1}{n}}, \quad (2)$$

where $P$ and $\rho$ denote the pressure and density, respectively. Equation (2) can be modified as

$$\frac{P}{Y} = \frac{K'}{Y}\left(\frac{\rho_{bulk}}{\delta}\right)^{\frac{n+1}{n}} = \alpha f^{\frac{n+1}{n}}, \quad (3)$$

where $K'$ and $\alpha$ are constants. We approximated the compression curves in Fig. 2 with Equation (3) and then obtained the factor $K'$, $\alpha$ and polytropic index $n$ for each sample. The results are listed in Table 2. We calculated $K'$ and $n$ from the data in our previous work (Omura et al., 2016; Omura and Nakamura 2017) and Güttler et al. (2009) using Equation (3). The pressure range of the dataset used for fitting was chosen because the data fitted the power law well. We also calculated $n$ from the data of Castellanos et al. (2005). In this case, the pressure of the original data was normalized; therefore, we show only the value of $n$. These values are listed in Table 2.

Previous studies also used a power law to describe the compression curve of granular material. For example, Kataoka et al. (2013a) gave the following relationship between pressure and the filling factor of highly porous granular aggregate:

$$P = \frac{E_{roll}}{{r_0}^3} f^3, \quad (4)$$

where P is the pressure, $r_0$ is the radius of the constituent particles, $f$ is the filling factor of the granular aggregate, and $E_{\mathrm{roll}}$ is the rolling energy of the constituent particles (see Wada et al. 2007). In this equation, pressure is proportional to the cube of the filling factor. Sirono (2004) fitted compressive strength data obtained experimentally using mixtures of toner particles and different amounts of fumed silica particles with the power law $\Sigma(f) = \Sigma_0 f^{\beta}$, where $\Sigma(f)$ is the compressive strength of a sample with a given filling factor $f$. $\Sigma_0$ and β are fitting parameters. The power exponent β corresponds to $(n + 1)/n$ in Eq. (1). We calculated $n$ from the power exponent value shown in Kataoka et al. (2013a) and Sirono (2004) and these are listed in Table 2. The values of $n$ obtained from the compression curves measured in our study ranged from ~0.02 to ~0.2. The value obtained from the compression curves of silica spheres (Güttler et al. 2009) was 0.34, which was slightly larger than our results. The values obtained from a compression curves of mixture of toner particles and fumed silica (Sirono 2004; Castellanos et al., 2005) or $TiO_2$ particles (Castellanos et al. 2005), and the samples used in our previous studies, were also within this range, except for the values of the 77 μm alumina particles. The value of $n$ obtained from the simulation (Kataoka et al. 2013a), which started with a more porous aggregate ($f \cong 10^{-3}$) than the samples used in this study, was 0.5, which was larger than the experimental results.

Figures 3a and 3b show the relationships between $n$ and the ranges of the filling factor and particle diameter, respectively. For the toner particles with additives, we adopted the diameter of the toner particles as the particle diameter. For alumina (0.1 μm), we adopted both the manufacturer's diameter and the median diameter obtained by measurement. In Figure 3b, both are shown in different symbols. Although the range of filling factors for silica beads and sand grains is similar, the value of $n$ varies by an order of magnitude and decreases with increasing particle diameter. This occurs because when the sample consists of smaller particles, there are more contact points, i.e., movable points, in the same volume. Particles can slide or rotate at contact points. This is probably why compression of a sample consisting of small particles is easier than compression of a sample consisting of large particles. For alumina (0.1 μm), the value of $n$ is similar to that of silica beads (1.7 μm) and the median diameter obtained by measurement is close to the median diameter of silica beads (1.7 μm). For the alumina sample, compression seems to be caused by rearrangement of aggregates prior to rearrangement of individual particles.

A popular formula for compression curves used in pharmacology is the Kawakita equation (Kawakita & Lüdde 1971)

$$\frac{P_a}{C} = \frac{1}{ab} + \frac{P_a}{a}, \quad (5)$$

where $P_a$ is the applied axial pressure and $C$ is the relative volume decrease, or

$$C = \left(\frac{V_0 - V}{V_0}\right) = \left(\frac{f - f_i}{f}\right) \quad (6)$$

where $V_0$ is the initial sample volume, $V$ is the sample volume, $f_i$ is the initial filling factor of the sample, and $a$ and $b$ are constants. Note that $C$ corresponds to a strain in the geometry of piston compression.

The Nutting equation is another equation that involves the relative volume decrease as a parameter (Nutting 1921; Scott Blair & Caffyn 1949)

$$\gamma = \psi^{-1} \sigma^{\mu} t^{\nu} \quad (7),$$

where $\gamma$ is the strain, $\sigma$ is the stress, t is the duration of the stress, $\mu$ and $\nu$ are physical exponents that change with the particle diameter and water content (Taneya & Sone 1962), and $\psi$ is a physical property called "firmness" (Scott Blair & Valda Coppen 1940). For $\mu = 1$ and $\nu = 1$, $\psi$ corresponds to the viscosity, while in the case of $\mu = 1$ and $\nu = 0$, $\psi$ corresponds to the elastic modulus (Nutting 1921). Here, $\gamma$ and $\sigma$ in Equation (7) are rewritten using $C$ and $P_a$, respectively, and Equation (7) is modified as

$$\frac{P_a}{C} = \psi {P_a}^{1-\mu} t^{-\nu} \quad (8).$$

In our experiments, the loading velocity was fixed. Therefore, we regarded $t^{-\nu}$ as a constant.

We fitted our experimental data with the Kawakita and Nutting equations. The results are shown in Fig. 4. Our experimental data were fitted well by both equations; however, the fit with the power-law form shown above was as good as the fittings with these equations. The correlation coefficients for all of the fittings exceeded 0.99.

3. INTERNAL POROSITY STRUCTURE OF GRANULAR OBJECTS

We calculated the internal structure of a granular body with a certain radius using the power-law relationships between pressure and the filling factor (i.e., density) determined by the measurements given in the previous section and the Lane–Emden equation:

$$\frac{1}{\xi^2}\frac{d}{d\xi}\left(\xi^2\frac{d\Phi}{d\xi}\right) = -\Phi^n. \quad (9)$$

$\Phi$ denotes a dimensionless parameter introduced for rewriting the density $\rho$ using the density of the center $\rho_c$ as follows:

$$\rho = \rho_c \Phi^n, \quad (10)$$

and $\xi$ denotes a dimensionless form of the radius of the object and is given by

$$\xi = \frac{r}{a}, \quad (11)$$

where

$$a = \sqrt{\frac{(n+1)K{\rho_c}^{\frac{1-n}{n}}}{4\pi G}}. \quad (12)$$

Here, $G$ is the gravitational constant.

The boundary conditions of the Lane–Emden equation are $\Phi = 1$ and $\frac{d\Phi}{d\xi} = 0$ at the center of the body. When we choose a set of constants $K$ and $n$ obtained from measurements of the compression properties of a granular sample, we can obtain the relationship between the radius $R$ and mass $M$ of a spherical body. Equation (9) was solved using the fourth-order Runge–Kutta method. Figure 5 shows the results for bodies with radii of 1, 10, 60, and 100 km when using the compression properties of the silica beads (1.7 μm) and silica sand (13 μm). Density was converted into porosity using $\varepsilon = 1$ $-(\rho_{bulk}/\delta)$, where $\varepsilon$ is the porosity. The porosity decreases with increasing distance from the surface and the radius of the object. The porosity structure depends on the compression properties of the constituent granular particles. In case of the silica beads, the porosity dependence on the depth from the surface of the body is stronger than that of the silica sand. The bulk porosity depends on the size of body more strongly for silica beads than the silica sand. The bulk porosity is almost equal for bodies 60 km in radius but consisting of different particles, i.e., silica beads and silica sand, although the radial porosity structure is different.

The results calculated in this study provide the porosity structure arising from soil pressure due only to self-gravity. That is, these are the initial, most porous structures of bodies consisting of the assumed granular materials when the centrifugal and tidal forces are negligibly small. Therefore, we can calculate the porosity structure of a planetesimal and a re-accumulated rubble-pile body that consists of granular material with certain compression properties. In comparison, asteroids lose their void spaces during their lifetimes, e.g., by impact compression or by compaction due to vibration, or because the bodies may contain rocks inside them. In these cases, we can constrain the internal structure of the body when we assume the compression properties of the constituent granular material. For example, 702 Alauda, which has a radius of ~100 km (Tedesco et al. 2002), has a bulk porosity calculated from its bulk density (Rojo & Margot 2011) and the density of CM chondrites (Consolmagno et al. 2008) of ~0.46. This value is smaller than the result calculated using the compression properties of silica sand shown in Fig. 5. If we assume that 702 Alauda consists of granular material that has the same compression properties as silica sand, this body contains rocks or has lost its void spaces during its lifetime.

In this study, we used compression curves measured in the laboratory. The compression properties of a granular material should depend on the interparticle force of its constituent particles, which varies with the measurement environment due to adsorbed molecules under ambient conditions (Perko et al. 2001; Kimura et al. 2015). The interparticle force increases when adsorbed molecules are removed; the compression of granular material may become more difficult in space compared to in a laboratory setting. In addition, the internal structure near the surface of a body was not constrained in this study because under a pressure smaller than Yield strength, the layer is not compressed to large degree and so the porosity of the layer near the surface may have remained essentially constant (Omura & Nakamura 2017).

## 4. SUMMARY

In this study, we conducted compression experiments using fluffy granular samples with various particle sizes, shapes, and compositions. We obtained a relationship between the applied pressure and filling factor for each sample and approximated it with a power-law form, i.e., a modified polytropic relationship. The approximation with a power-law form was as good as the approximation with the Kawakita equation, which is used in pharmacology. The polytropic index *n* for our samples ranged from ~0.02 to ~0.2 and decreased with the size of constituent particles. The Lane–Emden equation was applied and we presented a model of the radial porosity structures of granular bodies with different radii based on the measured compression curves. Different porosity structures were estimated based on different compression curves for bodies with the same size and bulk porosity. The results provide the initial, most porous structures of accreted primordial bodies or re-accumulated rubble-pile bodies when the centrifugal and tidal forces are negligibly small. The internal structure of small bodies can be constrained by comparing the results with the values based on observations.

We thank M. Hyodo for allowing us access to the laser diffractometer. We are grateful to an anonymous reviewer for constructive comments on this paper. This research was supported by the Hosokawa Powder Technology Foundation and a grant-in-aid for scientific research from the Japanese Society for the Promotion of Science of the Japanese Ministry of Education, Culture, Sports, Science, and Technology (MEXT; No. 25400453).

Table 1 Properties of the sample powders.

| Name | Grain Density δ ($gcm^{-3}$) | Shape | Material | Median Diameter (μm) | $d_{85}/d_{15}$ |
|---|---|---|---|---|---|
| **Alumina (0.1 μm)** | 3.9 | irregular | $Al_2O_3$ | ＜0.1[a], 2.6[b] | 2.9 |
| **Silica beads (1.7 μm)** | 2.2 | spherical | $SiO_2$ | 1.7[b] | 2.8 |
| **Silica sand (13 μm)** | 2.645 | irregular | $SiO_2$ | 13[b] | 12 |
| **Silica sand (19 μm)** | 2.645 | irregular | $SiO_2$ | 19[b] | 22 |
| **Silica sand (73 μm)** | 2.645 | irregular | $SiO_2$ | 73[b] | 2.0 |

[a] Manufacturer’s information, Goodfellow.

[b] Determined from the measured particle size distribution.

Table 2 Fitting parameters for the compression curves.

| | K' (Pa) | α | n | Ref. |
|---|---|---|---|---|
| **Alumina (0.1 µm)** | $(1.213\pm0.035)\times10^{11}$ | $(4.33\pm0.13)\times10^{2}$ | $(2.3164\pm0.0056)\times10^{-1}$ | … |
| **Silica beads (1.7 µm)** | $(8.35\pm0.23)\times10^{6}$ | $(1.391\pm0.039)\times10^{-1}$ | $(2.201\pm0.013)\times10^{-1}$ | … |
| **Silica sand (13 µm)** | $(4.96\pm0.49)\times10^{12}$ | $(8.27\pm0.81)\times10^{4}$ | $(4.437\pm0.025)\times10^{-2}$ | … |
| **Silica sand (19 µm)** | $(6.32\pm0.45)\times10^{11}$ | $(1.056\pm0.075)\times10^{4}$ | $(4.244\pm0.019)\times10^{-2}$ | … |
| **Silica sand (73 µm)** | $(4.22\pm0.36)\times10^{21}$ | $(7.05\pm0.60)\times10^{13}$ | $(2.0111\pm0.0045)\times10^{-2}$ | … |
| **Alumina (4.5 µm)** | $(1.59\pm0.12)\times10^{13}$ | … | $(5.647\pm0.023)\times10^{-2}$ | 1 |
| **Alumina (23 µm)** | $(4.50\pm0.80)\times10^{20}$ | … | $(2.521\pm0.013)\times10^{-2}$ | 1 |
| **Alumina (77 µm)** | $(1.24\pm0.49)\times10^{41}$ | … | $(8.430\pm0.041)\times10^{-3}$ | 1 |
| **Fly ash (4.8 µm)** | $(1.371\pm0.042)\times10^{9}$ | … | $(6.844\pm0.023)\times10^{-2}$ | 1 |
| **Alumina (6.5 µm)** | $(7.36\pm0.54)\times10^{11}$ | … | $(5.463\pm0.023)\times10^{-2}$ | 2 |
| **Alumina (15 µm)** | $(3.13\pm0.22)\times10^{12}$ | … | $(4.761\pm0.018)\times10^{-2}$ | 2 |
| **Alumina (23 µm)** | $(6.9\pm1.3)\times10^{19}$ | … | $(2.559\pm0.014)\times10^{-2}$ | 2 |
| **Fly ash (4.8 µm)** | $(1.649\pm0.025)\times10^{8}$ | … | $(8.134\pm0.014)\times10^{-2}$ | 2 |
| **Glass beads (18 µm)** | $(2.951\pm0.093)\times10^{16}$ | … | $(2.3496\pm0.0027)\times10^{-2}$ | 2 |
| **Silica sphere (1.5 µm)** | $(8.6\pm3.9)\times10^{5}$ | … | $(3.420\pm0.043)\times10^{-1}$ | 3 |
| **Toner (12.7 µm) + Silica (0.01 %)** | $1.9\times10^{6}$ | … | 0.152 | 4 |
| **Toner (12.7 µm) + Silica (0.05 %)** | $1.5\times10^{5}$ | … | 0.192 | 4 |
| **Toner (12.7 µm) + Silica (0.1 %)** | $3.3\times10^{5}$ | … | 0.159 | 4 |
| **Toner (12.7 µm) + Silica (0.2 %)** | $2.6\times10^{5}$ | … | 0.152 | 4 |
| **Toner (12.7 µm) + Silica (0.4 %)** | $3.3\times10^{6}$ | … | 0.105 | 4 |
| **Toner (7.8 μm) + $TiO_2$ (100 % SAC)** | … | … | $(6.47\pm0.12)\times10^{-2}$ | 5 |
| **Toner (7.8 μm) + Silica (32 % SAC)** | … | … | $(5.93\pm0.21)\times10^{-2}$ | 5 |
| **Toner (11.8 μm) + Silica (32 % SAC)** | … | … | $(5.51\pm0.19)\times10^{-2}$ | 5 |
| **Toner (15.4 μm) + Silica (32 % SAC)** | … | … | $(5.32\pm0.15)\times10^{-2}$ | 5 |
| **Toner (19.1 μm) + Silica (32 % SAC)** | … | … | $(4.96\pm0.16)\times10^{-2}$ | 5 |
| **Numerical simulation** | $4.7\times10^5$ (ice 0.2 μm)<br>$6.6\times10^3$ (silica 1.2 μm) | … | 0.50 | 6 |

**References.** (1) Omura et al. (2016), (2) Omura & Nakamura (2017), (3) Güttler et al. (2009), (4) Sirono (2004), (5) Castellanos et al. (2005), (6) Kataoka et al. (2013a).

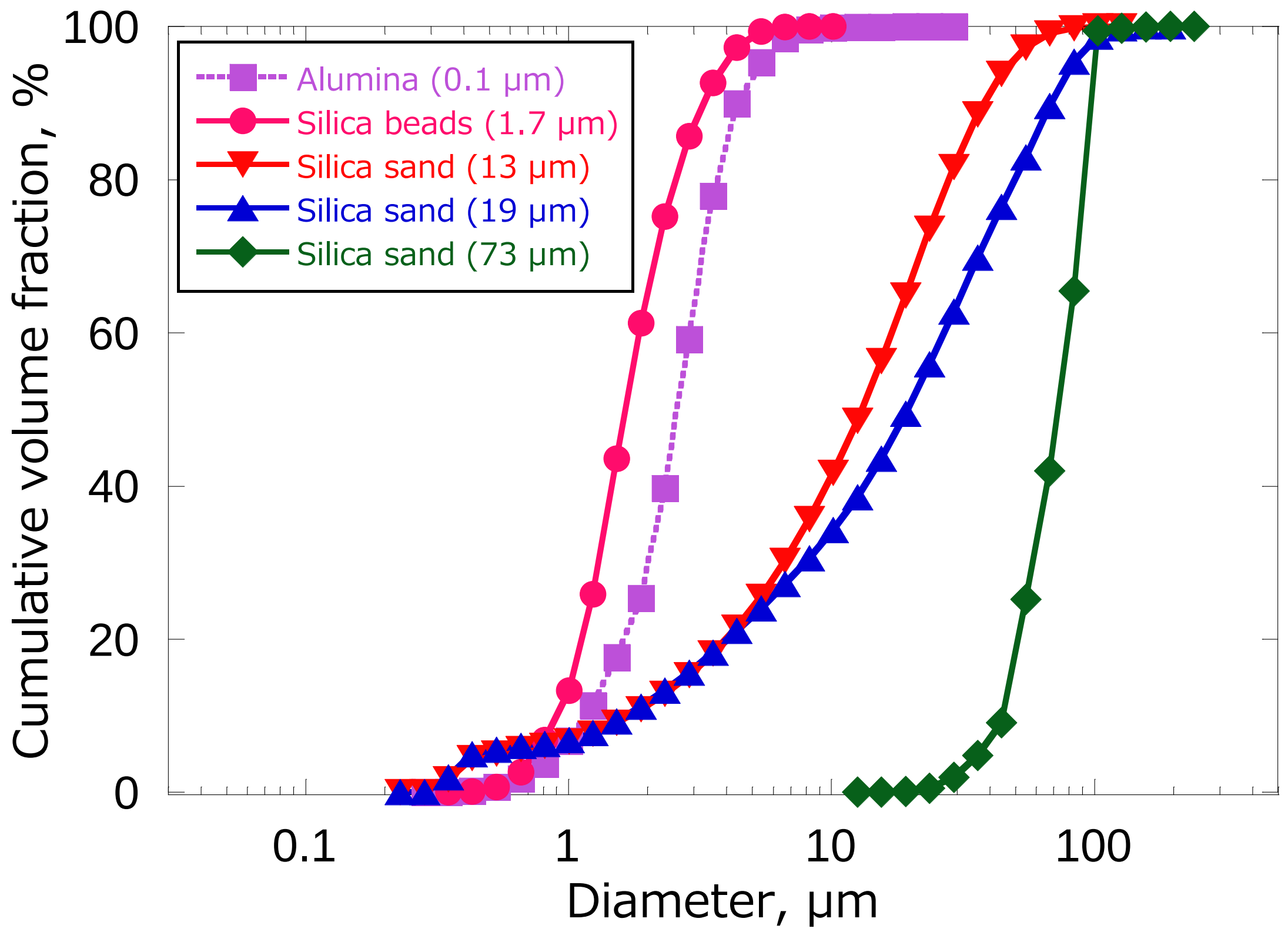

Fig. 1. Particle size distribution of the sample powders.

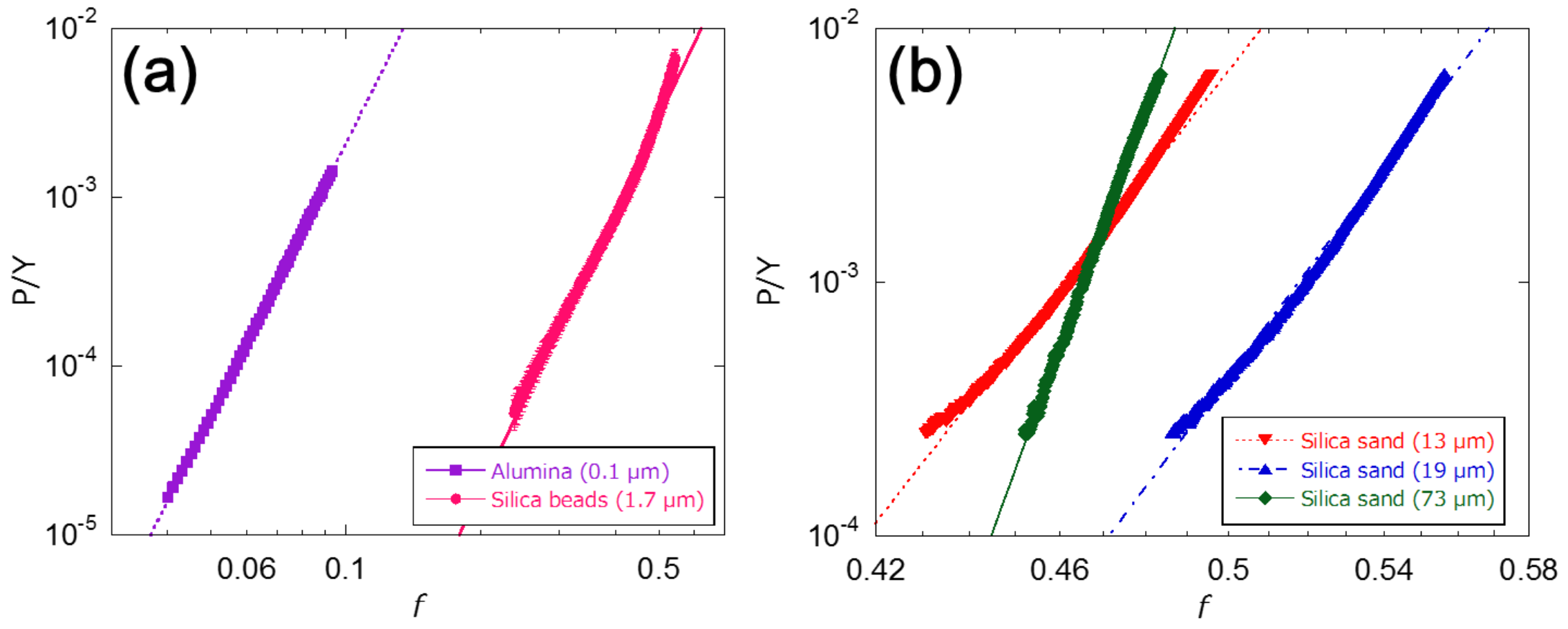

Fig.2. Relationships between the filling factor ($f$) and the applied pressure normalized by the tensile strength of the sample particles ($P/Y$). The thin lines are fitted according to Equation (3).

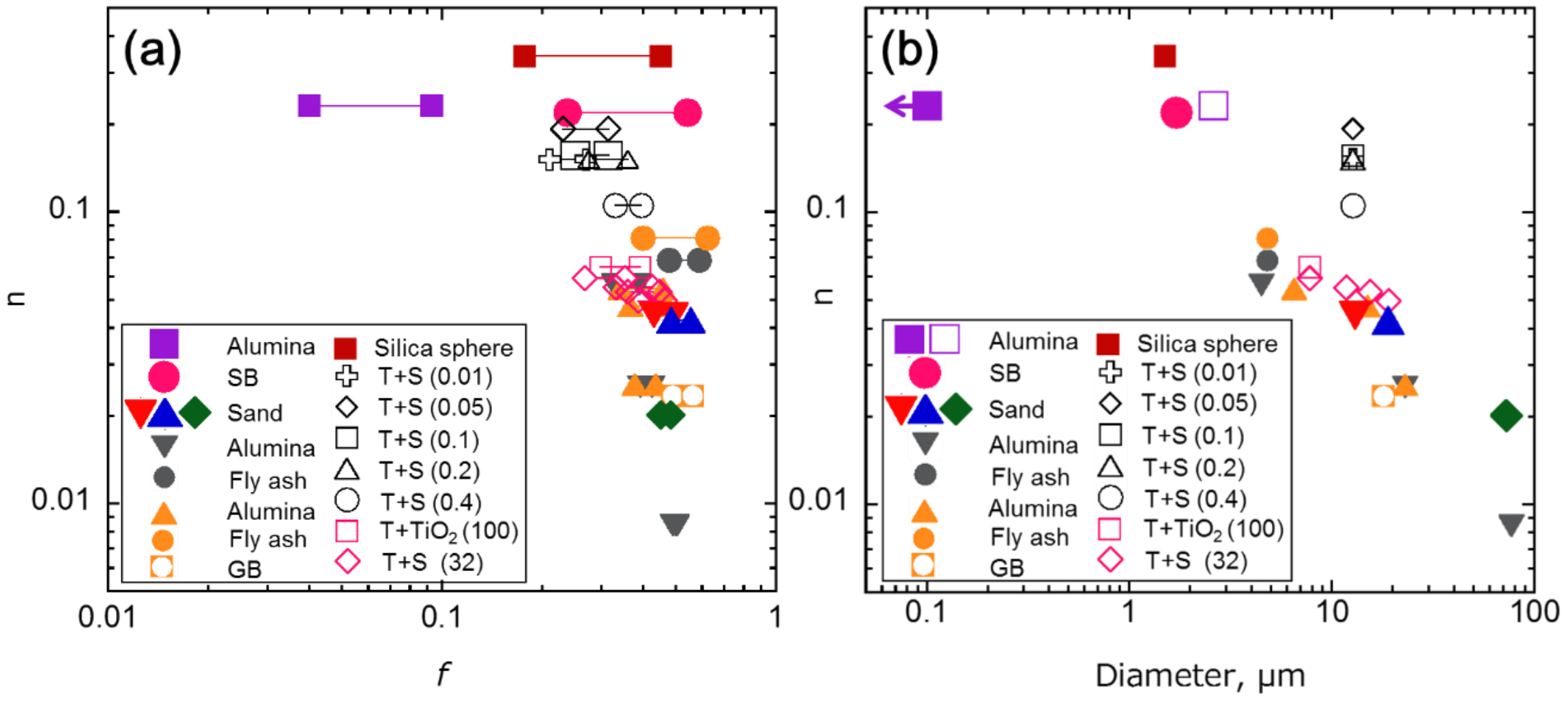

Fig. 3. Fig. 3. Relationship between the polytropic index (*n*) and (a) filling factor (*f*) and (b) particle diameter. Purple squares indicate the values of alumina (0.1 μm) and an open purple square is the median diameter obtained by measurement (see Table 1). SB indicates silica beads. Gray and orange symbols indicate the values obtained from Omura et al. (2016) and Omura & Nakamura (2017), respectively. GB indicates glass beads. The value indicated by a brown square was obtained from Güttler et al. (2009). Open black symbols show the values obtained by Sirono (2004). T+S indicates toner + fumed silica particles (7 nm) and values in parentheses are the concentrations of the additives in mass (%). Values calculated with the data obtained by Castellanos et al. (2005) are denoted by open pink symbols. T+$TiO_2$/S indicates toner + $TiO_2$ or fumed silica particles (8 or 50 nm) and the values in parentheses are the surface area coverage (SAC) of toner particles by additives (%). (For interpretation of the references to color in this figure legend, the reader is referred to the web version of this article.)

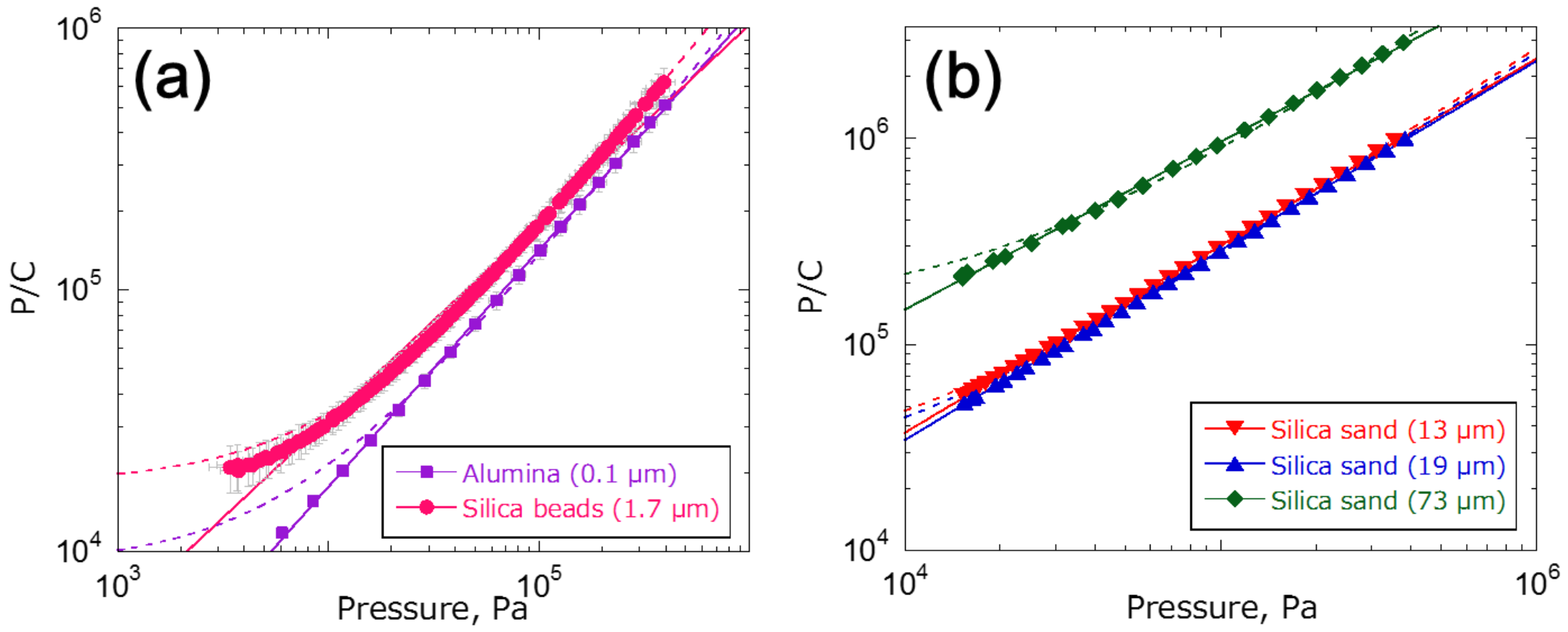

Fig. 4. Experimental results (marks) fitted with the Kawakita equation (thin dotted lines) and the Nutting equation (thin solid lines). To make the lines clearer, only 33 % (a) or 10 % (b) of the data points are plotted.

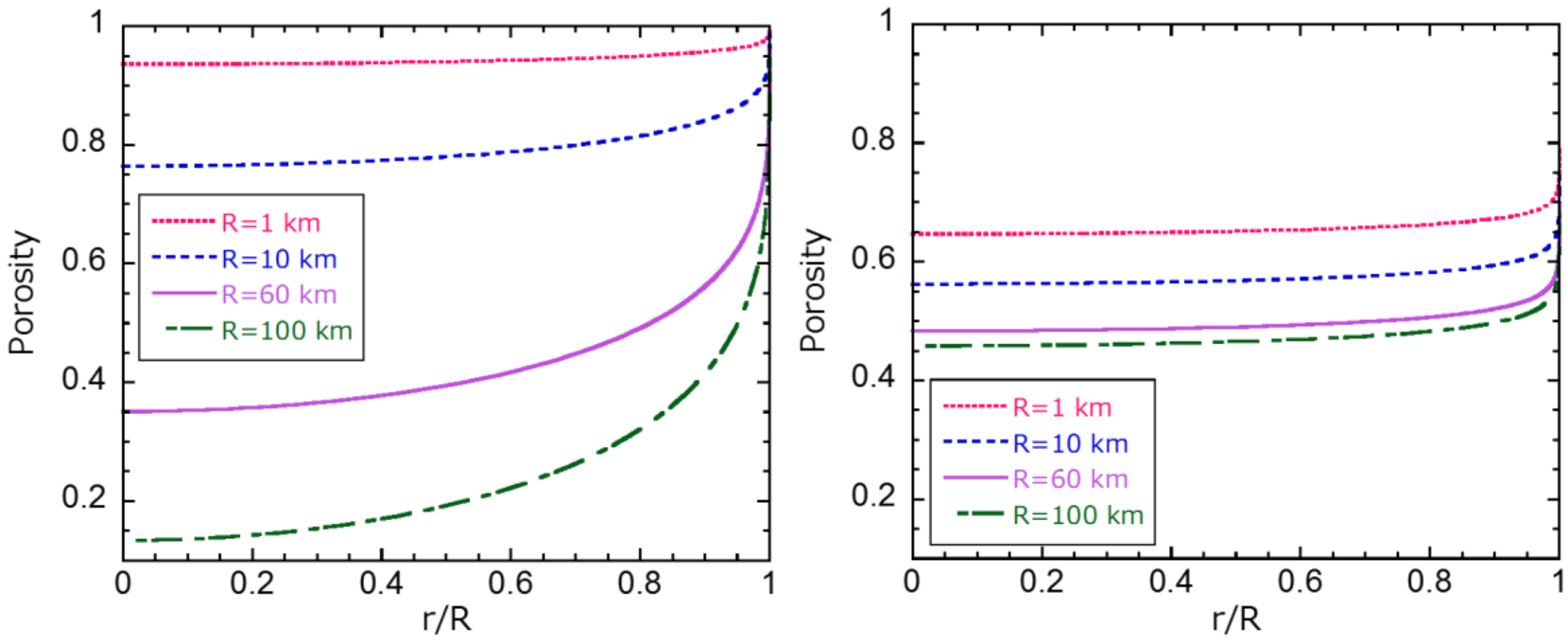

Fig. 5. Calculation results of the internal porosity structure of four different sized bodies. The horizontal axis shows the distance from the center normalized by the radius of the body. Curves show the results of the calculation. Left: calculated results using the compression properties of silica beads ($K = 2.48 \times 10^{-12}$, $n = 0.2201$). The bulk porosity is 0.95, 0.82, 0.51, and 0.34 for bodies of radius 1, 10, 60, and 100 km, respectively. Right: calculated results using the compression properties of 13 μm silica sand ($K = 1.4 \times 10^{-68}$, $n = 0.04437$). The bulk porosities are 0.67, 0.59, 0.51, and 0.49 for bodies of radius 1, 10, 60, and 100 km, respectively.